\documentclass[graybox]{svmult}

\usepackage{type1cm}        
\usepackage{makeidx}         
\usepackage{graphicx}        
\usepackage{multicol}        
\usepackage[bottom]{footmisc}
\usepackage{subfig}
\usepackage[font=small,labelfont=bf]{caption}

\usepackage{newtxtext}       %
\usepackage[varvw]{newtxmath}       

\makeindex             

\begin{document}

\title*{Adaptation Strategy for a Distributed Autonomous UAV Formation in Case of Aircraft Loss}
\author{Tagir Muslimov}
\institute{Tagir Muslimov \at Ufa State Aviation Technical University, Ufa, Russia, \email{tagir.muslimov@gmail.com}
}
%
%
\titlerunning{Adaptation Strategy for a Distributed Autonomous UAV Formation...}
\maketitle

\abstract{Controlling a distributed autonomous unmanned aerial vehicle (UAV) formation is usually considered in the context of recovering the connectivity graph should a single UAV agent be lost. At the same time, little focus is made on how such loss affects the dynamics of the formation as a system. To compensate for the negative effects, we propose an adaptation algorithm that reduces the increasing interaction between the UAV agents that remain in the formation. This algorithm enables the autonomous system to adjust to the new equilibrium state. The algorithm has been tested by computer simulation on full nonlinear UAV models. Simulation results prove the negative effect (the increased final cruising speed of the formation) to be completely eliminated.}


\section{Introduction}
\label{sec:1}
For a decentralized formation of unmanned aerial vehicles (UAVs), the formation needs to be capable of reconfiguration should one of the UAVs fail. Some papers approach this problem by developing algorithms to optimize the recovery or maintenance of connectivity. For instance, paper [1] investigates reconfiguring the communication topology to minimize the communication cost in case of the UAVs in the formation losing their links. When considering a UAV formation as a multiagent system, existing approaches become applicable, see, e.g., [2] for an overview. Paper [3] investigates the possibility of reaching consensus should agents stop communicating. Work [4] covers the effectiveness of a swarm intelligence in a multiagent system that loses its agents. Article [5] investigates controlling a multiagent system when links between agents are lost. Paper [6] proposes a novel approach that relies on a Negative Imaginary (NI) theory-based control algorithm that keeps a decentralized system of single integrators stable when several agents are lost. Work [7] covers an agent loss-tolerant multiagent system. One of the approaches covered therein consists in adding redundant links to make the group robust to agent loss. 

Being a complex system, a UAV formation does affect the reconfiguration algorithms in development. The way it does so depends, for instance, on the mission being planned or on the dynamics of the aircraft themselves. Paper [8] presents a concept of fault-tolerant formation control for cases of UAV loss. The control strategy consists in adjusting the objective function that keeps the entire formation functional. Work [9] covers a situation of UAV loss and proposes an algorithm that performs priority ranking in order to choose a UAV to replace the lost one. Article [10] presents an approach where the lost UAV agents are replaced on a priority basis in a formation caught in an natural disaster. Paper [11] covers operating a UAV formation in a wildfire. Should a single UAV be lost, the proposed algorithm reconfigures the formation by adjusting for the lost links. Numerous papers cover UAV formation control strategies in cases of actuator faults. For instance, work [12] demonstrates the use of UAV formations to tackle wildfires; it shows that cooperative tracking errors could be bounded even in case of actuator faults. Article [13] investigates UAV formation stability by applying a Lyapunov function, assuming both actuator faults and communication interrupts. Adaptive control enabled the authors to compensate for actuator faults.

The above paper overview leads to a conclusion that the existing publications make little focus on how losing a UAV agent affects the dynamics of the entire formation. Of particular interest are cases where the system has completely decentralized control, i.e., each UAV agent relies exclusively on readings on the adjacent aircraft to stir itself. Yamaguchi et al. have developed one of the most effective control methods for distributed autonomous systems. Paper [14] was the first to propose a model that enabled completely decentralized control. The approach was further enhanced in a series of subsequent papers that showed using this method to enable a group of autonomous robots to generate a variety of shapes [15] as well as to encircle a target [16]. Yamaguchi state this approach is based on formation vectors. The idea is that each agent in a formation has its own formation vector that is calculated from the adjacent agents' readings. Work [17] also showed that when autonomous robotic agents are configured to maintain different distances to their neighbors, the general stability of the formation can be maintained by running a special adaptation algorithm.

However, this cycle of papers did not focus on losing an agent from the formation. Besides, they only covered linear integrators as agents. Paper [18] proposed an adaptation algorithm for a decentralized system of linear oscillators, which also covered the case of losing a single oscillator. This approach can be modified for use in a system of autonomous robots including flying ones/UAVs.

Thus, the novelty hereof lies in that we (i) investigate the feasibility of implementing an agent loss adaptation algorithm in a completely decentralized/distributed autonomous agents, from the standpoint of its dynamics; (ii) test the proposed algorithm on full nonlinear UAV models engaged in cooperative target tracking in a circular formation.

\section{Adaptation Algorithm for a Decentralized UAV Formation}

\subsection{Distributed Autonomous System Model}

%
\begin{figure}[b]
	\sidecaption
	\includegraphics[scale=.25]{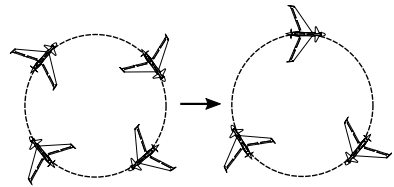}
	%
	%
	\caption{Losing one UAV in a formation due to failure}
	\label{fig:1}       
\end{figure}

\begin{figure}[b]
	\sidecaption
	\includegraphics[scale=.35]{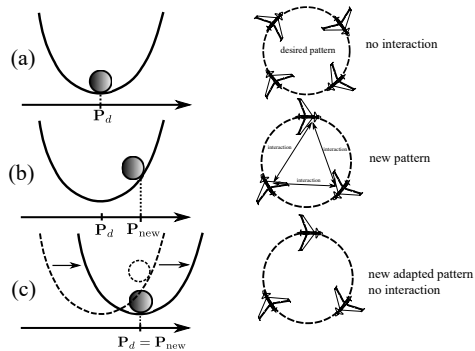}
	%
	%
	\caption{Reconfiguring a decentralized system by means of an adaptation algorithm. Figure partly adapted from [18]}
	\label{fig:2}       
\end{figure}

A decentralized system usually consists of interchangeable locally interconnected systems. The structural similarity of such systems to their biological counterparts is what makes researchers believe these systems could be flexible and adaptable. This gives rise to two problems: controlling the entire system with only local data at hand; and implementing adaptation mechanisms. Of particular interest is the development of a fault-tolerant algorithm to adapt the UAV formation to losing a faulty UAV agent when tracking a ground target. This model is based on the Japanese researcher's work [18–20], where they developed an adaptation algorithm for the locomotion model of an oscillatory network of motor neurons in a biological organism; they also developed a multicylinder engine model [21]. These papers consider locally interconnected oscillators that maintain relative phase angles to create locomotion rhythms that control walking, running, swimming, and other types of organism locomotion. In a target-tracking decentralized UAV formation, UAV agents, too, communicate locally and maintain the present relative phase angles (phase shifts) of orbiting the target. This is why this model, when further modified and developed, could become the foundation for an adaptive UAV formation reconfiguration algorithm for fault-tolerant control.

We further describe a distributed autonomous system per [18]. The system consists of $N$ interchangeable subsystems ${\left\{ {{S_i}} \right\}_{i = 1, \ldots ,m}}$. Interchangeability means that the dynamics of such subsystems can be described with differential equations of the same form. For analysis, consider a subsystem state written as the state variable ${q_i}$. This could be the phase angle of orbiting the target in a formation. Consider the vector ${\bf{Q}} = {\left[ {{q_1}, \ldots ,{q_m}} \right]^T}$. 

In a decentralized formation, UAVs do not have one-to-all communications, i.e., a single UAV agent will not be able to communicate with all other UAV agents in the formation. In other words, the UAV agent dynamics contains the state variables for the limited set of adjacent UAV agents. Interaction refers herein to receiving the state readings of the adjacent UAV agents. Let $N\left( {{q_i}} \right) = \left\{ {{q_l}|{q_l}\;{\rm{is\_interacting \_with}}\;{q_i}} \right\}$ be the set of such adjacent UAVs.

Each pair of interacting UAVs, e.g., ${S_i}$ and ${S_j}$, have a linear functional dependency relationship as ${p_k} = {p_k}\left( {{q_i},{q_j}} \right)$, where $k = 1, \ldots ,n$ are the ordinal numbers of interactions. The difference between the respective subsystem states is an example of such functional dependency. Consider the vector ${\bf{P}} = {\left[ {{p_1}, \ldots ,{p_n}} \right]^T}$. The functional dependencies in the formation can be written as the equation ${\bf{P}} = {\bf{LQ}}$, where ${\bf{L}} \in {R^{n \times m}}$. 

Control consists in finding such subsystem dynamics ${\dot q_i} = {f_i}\left( {N\left( {{q_i}} \right)} \right)$ that the functional dependencies ${\bf{P}}$ converge to the desired values ${{\bf{P}}_d} = {\left[ {{p_{d1}}, \ldots ,{p_{dn}}} \right]^T}$. For brevity, let us refer to the desired geometric shape of the UAV formation (depends on ${{\bf{P}}_d}$) as the pattern. An important assumption here is that the desired values of the functional dependencies ${{\bf{P}}_d}$ are predetermined, i.e., there is such vector ${\bf{Q}}$ that the condition ${\bf{LQ}} = {{\bf{P}}_d}$ holds. Write such condition as $\left( {{\bf{I}} - {\bf{L}}{{\bf{L}}^ + }} \right){{\bf{P}}_d} = {\bf{0}}$, where ${{\bf{L}}^ + }$ is the pseudoinverse of matrix ${\bf{L}}$, and ${\bf{I}}$ is a unit matrix. 

UAV formation control applies when the vector  ${\bf{P}}$ converges to the vector ${{\bf{P}}_d}$, i.e., each element of the vector ${\bf{P}}$ converges to the correspondingly numbered element of the vector ${{\bf{P}}_d}$. However, the vector ${{\bf{P}}_d}$ is not always attainable for a variety of reasons. For instance, it will not be attainable if one or more UAVs in a formation fail. Thus, creating an adaptation (or adaptive reconfiguration) algorithm boils down to developing a vector ${{\bf{P}}_d}$ tuning algorithm.

Yuasa and Ito published a series of papers [19, 20] where they solved the above-mentioned control problem on determining the dynamics of subsystems ${\dot q_i} = {f_i}\left( {N\left( {{q_i}} \right)} \right)$ for oscillators by proving a number of theorems. However, their solution was developed with linear agents in mind and would require a substantial modification for use on nonlinear agents. To begin with, the equation  ${\dot q_i} = {f_i}\left( {N\left( {{q_i}} \right)} \right)$ can be rewritten in vector form as $\dot {\bf Q}  = {\bf{F}}\left( {\bf{Q}} \right),$ where ${\bf{F}} = {\left[ {{f_1}, \ldots ,{f_m}} \right]^T}$. The equation can be transformed as ${\dot{\bf P}} = {\bf{L}} \dot {\bf Q}$=${\bf{LF}}\left( {\bf{Q}} \right)$.

\begin{theorem} (Yuasa and Ito [19]). 
Dynamics of ${\bf{P}}$ in the equation ${\dot{\bf P}} = {\bf{L}} \dot {\bf Q}$=${\bf{LF}}\left( {\bf{Q}} \right)$ can be described as an autonomous system if and only if ${\bf{F}}$ in the equation $\dot {\bf Q}  = {\bf{F}}\left( {\bf{Q}} \right)$ satisfies the following condition: ${\bf{L}}\frac{{\partial {\bf{F}}}}{{\partial {\bf{Q}}}}\left( {{\bf{I}} - {{\bf{L}}^ + }{\bf{L}}} \right) = \bf 0$.
\end{theorem}

\begin{theorem} (Yuasa and Ito [19]). 
If the dynamics of ${\bf{P}}$ can be described as a gradient system with the potential function $V\left( {\bf{P}} \right)$, i.e., ${\bf{P}} =  - \frac{{\partial V}}{{\partial {\bf{P}}}}$, then ${\bf{F}}$ can be written as ${\bf{F}} = {\left( {\frac{{\partial {V_{\bf{X}}}\left( {\bf{X}} \right)}}{{\partial {\bf{X}}}}} \right)^T} + \left( {{\bf{I}} - {{\bf{L}}^ + }{\bf{L}}} \right){\bf{Q'}},$ where ${\bf{X}} =  - {{\bf{L}}^T}{\bf{P}}$, ${V_{\bf{X}}}\left( {\bf{X}} \right) = {V_{\bf{X}}}\left( { - {{\bf{L}}^T}{\bf{P}}} \right) = V\left( {\bf{P}} \right),$ and ${\bf{Q'}}$ is an arbitrary vector that has the same dimensionality as ${\bf{Q}}$. Conversely, if ${\bf{F}}$ can be described with the equation above, there exists such potential function $V\left( {\bf{P}} \right)$ that the dynamics of $\bf P$ would be described as ${\bf{P}} =  - \frac{{\partial V}}{{\partial {\bf{P}}}}$.
\end{theorem}

Suppose that the subsystem dynamics can be described as ${\dot q_i} = {\bar f_i} + {\tilde f_i}\left( {{x_i}} \right)$, where ${\left[ {{{\bar f}_1}, \ldots ,{{\bar f}_m}} \right]^T} = {\bf{\bar F}} \in \ker {\bf{L}}$ and ${\left[ {{{\tilde f}_1}, \ldots ,{{\tilde f}_m}} \right]^T} = {\bf{\tilde F}} \in {\left( {\ker {\bf{L}}} \right)^ \bot }$. Dynamics of $\bf{P}$ is only determined by the summand ${\tilde f_i}\left( {{x_i}} \right)$ since ${\bf{L\bar F}} = 0$. Given this, write the dynamics of $\bf{P}$ as $\dot {\bf P} = {\bf{L\tilde F}}$. Paper [18] also proves the following theorem. 

\begin{theorem} ([18]).
If ${\tilde f_i}\left( {{x_i}} \right)$ satisfies the following conditions:
1.	${\tilde f_i}\left( {{x_i}} \right) = 0$ at the point ${x_i} = {x_{di}}$, 2.${\left. {\frac{{\partial {{\tilde f}_i}\left( {{x_i}} \right)}}{{\partial {x_i}}}} \right|_{{x_i} = {x_{id}}}} > 0$,
then ${\bf{X}} = {{\bf{X}}_d}$ (here ${{\bf{X}}_d} =  - {{\bf{L}}^T}{{\bf{P}}_d}$) will be one of the equilibrium states. Besides, if the following holds: 3.${\tilde f_i}\left( {{x_i}} \right) \cdot \left( {{x_i} - {x_{di}}} \right) > 0,$ then ${\bf{X}} = {{\bf{X}}_d}$ becomes the only equilibrium state. 
\end{theorem}

The authors of [19] prove this theorem under an assumption that the dynamics of  $\bf{P}$ become a gradient system. When all the three conditions of this theorem are satisfied, ${\bf{X}} = {{\bf{X}}_d}$ is the global minimum of the potential function $V\left( {\bf{X}} \right) = \sum\nolimits_{i = 1}^m {\int {{{\tilde f}_i}\left( {{x_i}} \right)} } d{x_i}$. Below is a variant of dynamics satisfying this theorem:
\begin{equation}
	{\dot q_i} = {\bar f_i} + f_i^ + \left( {{x_i} - {x_{di}}} \right),
\end{equation}
where $f_i^ + \left( {\cdot} \right)$ is an odd increasing function.

\subsection{Adaptation Strategy for a Distributed Autonomous System}

In paper [18], the authors refer to losing one or more oscillators as an environment variation. In a UAV formation, it can be referred to as a structural variation in the formation. Should the formation lose a single UAV as shown in Fig. 1, the desired phase shifts between the UAVs will be altered, as they are assumed to depend on the total number of UAVs in the formation. In other words, the pattern ${{\bf{P}}_d}$ configured before losing a faulty UAV becomes unattainable. Pattern ${{\bf{P}}_d}$ being unattainable means there is no such matrix $\bf Q$ that would satisfy the equation ${\bf{LQ}} = {{\bf{P}}_d}$. From the above equations ${{\bf{X}}_d} =  - {{\bf{L}}^T}{{\bf{P}}_d}$ and ${\bf{X}} =  - {{\bf{L}}^T}{\bf{P}}$, derive ${\bf{X}} - {{\bf{X}}_d} =  - {{\bf{L}}^T}\left( {{\bf{P}} - {{\bf{P}}_d}} \right)$. Apparently, ${\bf{X}} - {{\bf{X}}_d} = \bf 0$ is necessary but insufficient to satisfy ${\bf{P}} - {{\bf{P}}_d} = \bf 0$. In other words, if ${\bf{P}} - {{\bf{P}}_d} \ne {\bf{0}}$ and also ${\bf{P}} - {{\bf{P}}_d} \in \ker {\bf{L}}$, then ${\bf{X}} - {{\bf{X}}_d} = \bf 0.$

Thus, ${{\bf{P}}_d}$ needs to be adaptively tuned. Such tuning involves estimating the current system. From the engineering standpoint, it is reasonable to use the function
\begin{equation}
E = \frac{1}{2}{\sum\nolimits_{i = 1}^n {\left( {{p_i} - {p_{di}}} \right)} ^2}.
\end{equation}
Fig. 2 shows the minimum of this function is attained when the desired pattern ${{\bf{P}}_d}$ is formed; local interactions between the adjacent UAVs are null. If the pattern ${{\bf{P}}_d}$ cannot be attained due to losing a UAV from the formation, then the interaction is ongoing, and a new pattern emerges as a result of balancing such interactions, see Fig. 2b. To reduce subsystem interaction while preserving the newly emerged pattern, ${{\bf{P}}_d}$ needs to be tuned, see Fig. 2c. 

As noted in [18], a UAV agent's  ${S_i}$ interaction as a subsystem can be defined in terms of the equation:
\begin{equation}
I_k^i =  - {L_{ki}}\left( {{p_k} - {p_{dk}}} \right).
\end{equation}
Here, $k$ is the ordinal number of the interaction, i.e., assuming that in the matrix $\bf{L}$, its elements ${L_{ki}}$ and ${L_{kj}}$ are non-zero, that means the UAVs numbered $i$ and $j$ interact as subsystems in the formation. 

Based on (3), the second term of the equation (1) can be written as [18]
\begin{equation*}
	f_i^ + \left( {{x_i} - {x_{di}}} \right) = f_i^ + \left( {\sum\nolimits_{k = 1}^n {I_k^i} } \right).
\end{equation*}

Using (3), the objective function (2) can be transformed as follows [18]:
\begin{equation*}
E = \frac{1}{4}\sum\limits_{i = 1}^m {\sum\limits_{\scriptstyle k = 1\hfill\atop
		\scriptstyle{L_{ki}} \ne 0\hfill}^n {\frac{1}{{L_{ki}^2}}} } {\left\{ {I_k^i} \right\}^2}.
\end{equation*}
Apparently, minimizing the function $E$, we minimize the interactions $I_k^i$ between the subsystems.

The key point of such adaptive reconfiguration is the separation of dynamics on the time scale, i.e., the system dynamics needs to be much faster than the adaptive tuning dynamics. Reason being, adaptation requires estimating the current state of the system. Before a pattern can be adjusted and a new variation thereof can be made, the current pattern (variation) needs to be suitability-tested. 

Let us show how this approach could be used on a decentralized UAV formation when tracking a target. Consider a 4-UAV formation orbiting the target at a preset radius. Control aims at maintaining a predefined phase shift between UAVs numbered $i$ and $i+1$. The UAVs are engaged in an "open-chain" interaction, i.e., 1-2, 2-3, 3-4. This phase shift is denoted as ${p_{i,i + 1}} = {\varphi _{i + 1}} - {\varphi _i}$, where ${\varphi _ * }$ is the phase angle of UAVs orbiting the target, with the subscript for the UAV's ordinal number in the formation. As this is a four-UAV group, $i = 1,2,3,4.$ Phase shifts should converge to the desired values determined by the vector of the pattern ${{\bf{P}}_d} = {\left[ {{D_1},{D_2},{D_3}} \right]^T} = {\left[ {{{2\pi } \mathord{\left/
				{\vphantom {{2\pi } {3,{{9\pi } \mathord{\left/
									{\vphantom {{9\pi } {13,}}} \right.
									\kern-\nulldelimiterspace} {13,}}{{18\pi } \mathord{\left/
									{\vphantom {{18\pi } {29}}} \right.
									\kern-\nulldelimiterspace} {29}}}}} \right.
				\kern-\nulldelimiterspace} {3,{{9\pi } \mathord{\left/
						{\vphantom {{9\pi } {13,}}} \right.
						\kern-\nulldelimiterspace} {13,}}{{18\pi } \mathord{\left/
						{\vphantom {{18\pi } {29}}} \right.
						\kern-\nulldelimiterspace} {29}}}}} \right]^T}$. Then we can obtain
\begin{equation*}
		{\bf{P}} = {\bf{LQ}},\;\;\;{\bf{L}} = \left[ {\begin{array}{*{20}{c}}
		{ - 1}&1&0&0\\
		0&{ - 1}&1&0\\
		0&0&{ - 1}&1
  		\end{array}} \right],
\end{equation*}
where ${\bf{P}} = {\left[ {\begin{array}{*{20}{c}}
			{{p_{12}}}&{{p_{23}}}&{p{}_{34}}
	\end{array}} \right]^T}$ and ${\bf{Q}} = {\left[ {\begin{array}{*{20}{c}}
			{{\varphi _1}}&{{\varphi _2}}&{{\varphi _3}}&{{\varphi _4}}
	\end{array}} \right]^T}.$ The kernel of the matrix ${\bf{L}}$ in this case is defined as $\ker {\bf{L}} = {\left[ {\begin{array}{*{20}{c}}
			1&1&1&1
	\end{array}} \right]^T},$ and from the relation ${\bf{X}} =  - {{\bf{L}}^T}{\bf{LQ}},$ it can be found that
\begin{equation*}
	{\bf{X}} = \left[ {\begin{array}{*{20}{c}}
			{{p_{12}}}\\
			{ - {p_{12}} + {p_{23}}}\\
			{ - {p_{23}} + {p_{34}}}\\
			{ - {p_{34}}}
	\end{array}} \right] = \left[ {\begin{array}{*{20}{c}}
			{ - {\varphi _1} + {\varphi _2}}\\
			{{\varphi _1} - 2{\varphi _2} + {\varphi _3}}\\
			{{\varphi _2} - 2{\varphi _3} + {\varphi _4}}\\
			{{\varphi _3} - {\varphi _4}}
	\end{array}} \right].
\end{equation*}

UAV formation dynamics in case of encircling a target is defined as
\begin{equation*}
{\dot{\bf Q}} = \left[ {\begin{array}{*{20}{c}}
		{{{\dot \varphi }_1}}\\
		{{{\dot \varphi }_2}}\\
		{{{\dot \varphi }_3}}\\
		{{{\dot \varphi }_4}}
\end{array}} \right] = \left[ {\begin{array}{*{20}{c}}
		{{\omega _1}}\\
		{{\omega _2}}\\
		{{\omega _3}}\\
		{{\omega _4}}
\end{array}} \right] + \left[ {\begin{array}{*{20}{c}}
		{{v_f}\left( {{2 \mathord{\left/
						{\vphantom {2 {\left\{ {\pi {\rho _1}} \right\}}}} \right.
						\kern-\nulldelimiterspace} {\left\{ {\pi {\rho _1}} \right\}}}} \right)\arctan \left( {{k_\theta }\left( {{p_{12}} - {D_1}} \right)} \right)}\\
		{{v_f}\left( {{2 \mathord{\left/
						{\vphantom {2 {\left\{ {\pi {\rho _2}} \right\}}}} \right.
						\kern-\nulldelimiterspace} {\left\{ {\pi {\rho _2}} \right\}}}} \right)\arctan \left( {{k_\theta }\left( { - {p_{12}} + {p_{23}} + {D_1} - {D_2}} \right)} \right)}\\
		{{v_f}\left( {{2 \mathord{\left/
						{\vphantom {2 {\left\{ {\pi {\rho _3}} \right\}}}} \right.
						\kern-\nulldelimiterspace} {\left\{ {\pi {\rho _3}} \right\}}}} \right)\arctan \left( {{k_\theta }\left( { - {p_{23}} + {p_{34}} + {D_2} - {D_3}} \right)} \right)}\\
		{{v_f}\left( {{2 \mathord{\left/
						{\vphantom {2 {\left\{ {\pi {\rho _3}} \right\}}}} \right.
						\kern-\nulldelimiterspace} {\left\{ {\pi {\rho _3}} \right\}}}} \right)\arctan \left( {{k_\theta }\left( { - {p_{34}} + {D_3}} \right)} \right)}
\end{array}} \right],
\end{equation*}
where ${\omega _ * }$ and ${\rho _ * }$ are the angular velocity of UAVs orbiting the target, and the radius of such orbit, respectively; subscripts match the UAVs' ordinal numbers in the formation. Applying a nonlinear $\arctan \left( {\cdot} \right)$ function limits the minima and maxima of the speed control signal. Thus, it avoids the undesirable integral windup that might cause the UAV formation to go unstable. Here ${v_f}$ is a positive constant that determines the maximum added speed; ${k_\theta }$ is a positive constant that determines the smoothness of how the UAV reaches its orbit.

The dynamics of the pattern ${\dot{\bf P}}$, once the uniform radius $\rho$ is attained, is written as follows since ${\omega _1} = {\omega _2} = {\omega _3} = {\omega _4} = {v \mathord{\left/
		{\vphantom {v \rho }} \right.
		\kern-\nulldelimiterspace} \rho },$ where $v$ is the final cruising linear speed of the formation:
\begin{equation*}
{\dot{\bf P}} = \left[ {\begin{array}{*{20}{c}}
		{{{\dot p}_{12}}}\\
		{{{\dot p}_{23}}}\\
		{{{\dot p}_{34}}}
\end{array}} \right] = \left[ {\begin{array}{*{20}{c}}
		\begin{array}{l}
			- {v_f}\left( {{2 \mathord{\left/
						{\vphantom {2 {\left\{ {\pi \rho } \right\}}}} \right.
						\kern-\nulldelimiterspace} {\left\{ {\pi \rho } \right\}}}} \right)\arctan \left( {{k_\theta }\left( {{p_{12}} - {D_1}} \right)} \right) + \\
			+ {v_f}\left( {{2 \mathord{\left/
						{\vphantom {2 {\left\{ {\pi \rho } \right\}}}} \right.
						\kern-\nulldelimiterspace} {\left\{ {\pi \rho } \right\}}}} \right)\arctan \left( {{k_\theta }\left( { - {p_{12}} + {p_{23}} + {D_1} - {D_2}} \right)} \right)
		\end{array}\\
		\begin{array}{l}
			- {v_f}\left( {{2 \mathord{\left/
						{\vphantom {2 {\left\{ {\pi \rho } \right\}}}} \right.
						\kern-\nulldelimiterspace} {\left\{ {\pi \rho } \right\}}}} \right)\arctan \left( {{k_\theta }\left( \begin{array}{l}
					- {p_{12}} + {p_{23}}\\
					+ {D_1} - {D_2}
				\end{array} \right)} \right) + \\
			+ {v_f}\left( {{2 \mathord{\left/
						{\vphantom {2 {\left\{ {\pi \rho } \right\}}}} \right.
						\kern-\nulldelimiterspace} {\left\{ {\pi \rho } \right\}}}} \right)\arctan \left( {{k_\theta }\left( \begin{array}{l}
					- {p_{23}} + {p_{34}}\\
					+ {D_2} - {D_3}
				\end{array} \right)} \right)
		\end{array}\\
		\begin{array}{c}
			- {v_f}\left( {{2 \mathord{\left/
						{\vphantom {2 {\left\{ {\pi \rho } \right\}}}} \right.
						\kern-\nulldelimiterspace} {\left\{ {\pi \rho } \right\}}}} \right)\arctan \left( {{k_\theta }\left( { - {p_{23}} + {p_{34}} + {D_2} - {D_3}} \right)} \right) + \\
			+ {v_f}\left( {{2 \mathord{\left/
						{\vphantom {2 {\left\{ {\pi \rho } \right\}}}} \right.
						\kern-\nulldelimiterspace} {\left\{ {\pi \rho } \right\}}}} \right)\arctan \left( {{k_\theta }\left( { - {p_{34}} + {D_3}} \right)} \right)
		\end{array}
\end{array}} \right].
\end{equation*}

We propose the following adaptation algorithm:
\begin{equation}
{\dot p_{di}} = {a_s}{f_{sigm}}\left( {{\tau _p}\left( {{p_i} - {p_{di}}} \right)} \right),
\end{equation}
where $0 < {\tau _p} < 1$; ${f_{sigm}}$ is a sigmoid odd function, e.g., an arctangent or hyperbolic tangent; ${a_s}$ is a positive coefficient to alter the maxima and minima of the function ${a_s}{f_{sigm}}$; for example, if an arctangent is picked for ${f_{sigm}}$ , then this can be ${a_s} = {2 \mathord{\left/
		{\vphantom {2 \pi }} \right.
		\kern-\nulldelimiterspace} \pi }$. 

\begin{figure}[b]
	
	\subfloat[]{\includegraphics[width=0.5\textwidth]{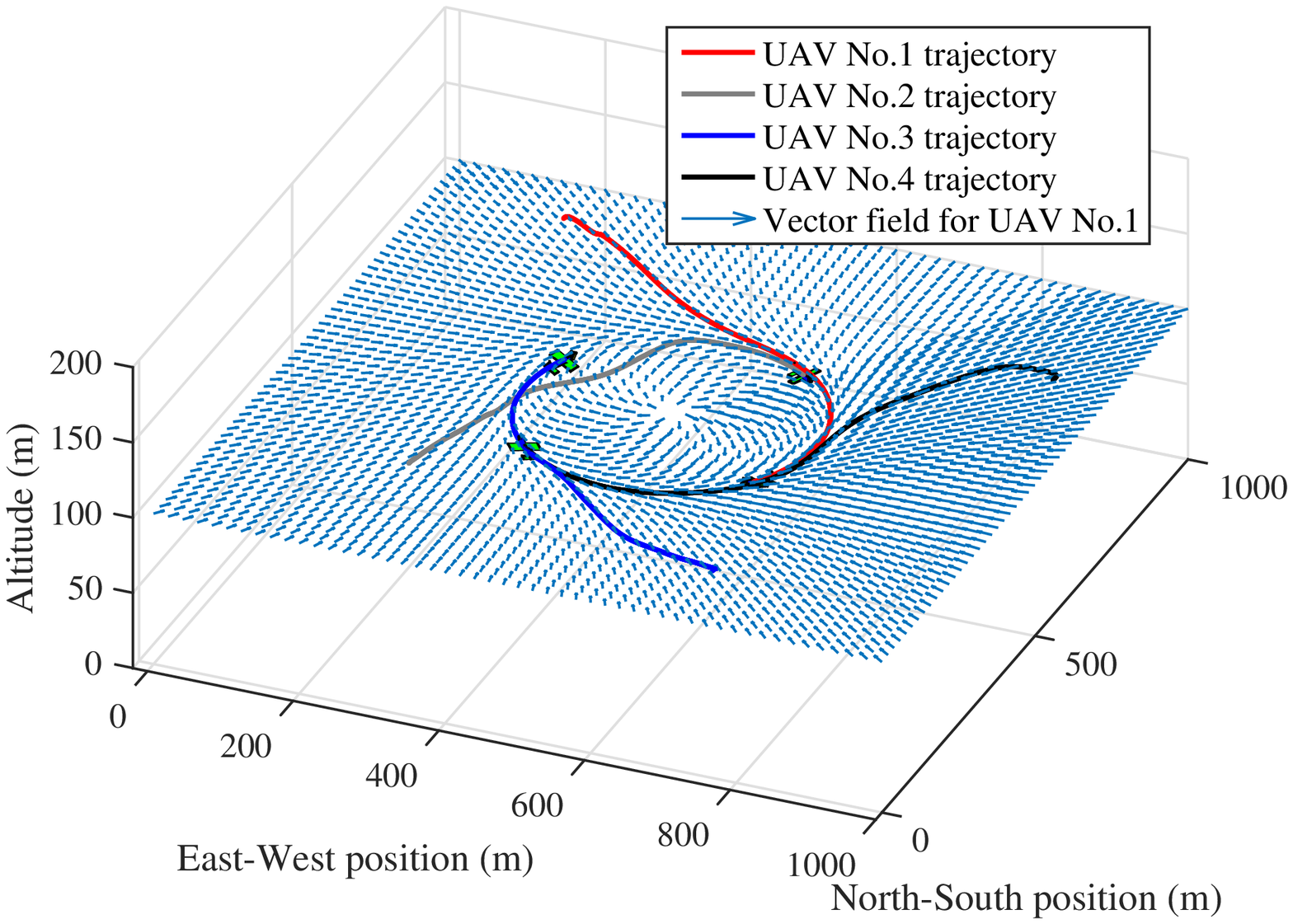}}
	\subfloat[]{\includegraphics[width=0.5\textwidth]{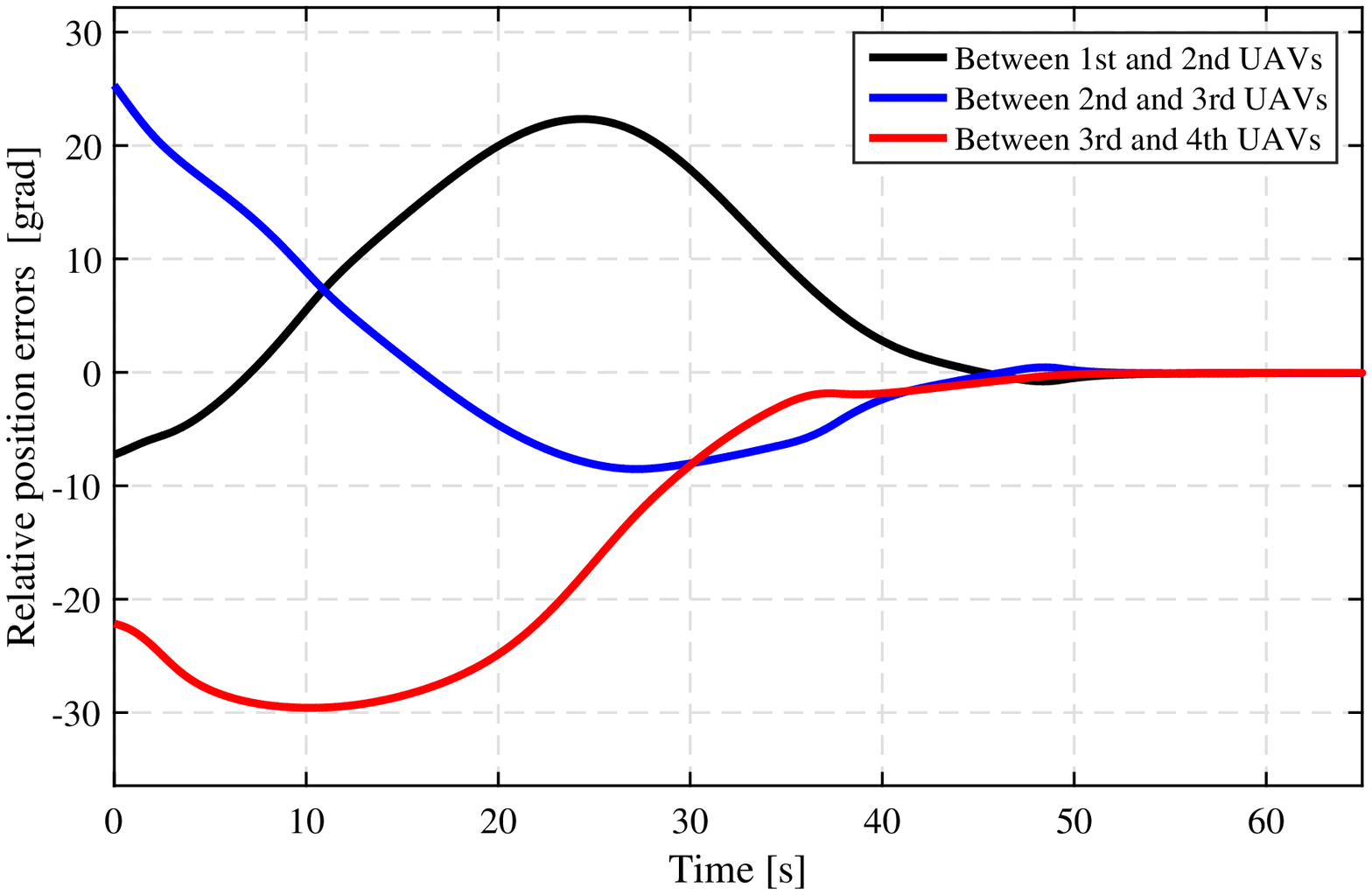}}
	
	\caption{(a) 4-UAV formation trajectories; (b) relative phase errors in a 4-UAV formation}
\end{figure}

\begin{figure}[b]
	
	\subfloat[]{\includegraphics[width=0.5\textwidth]{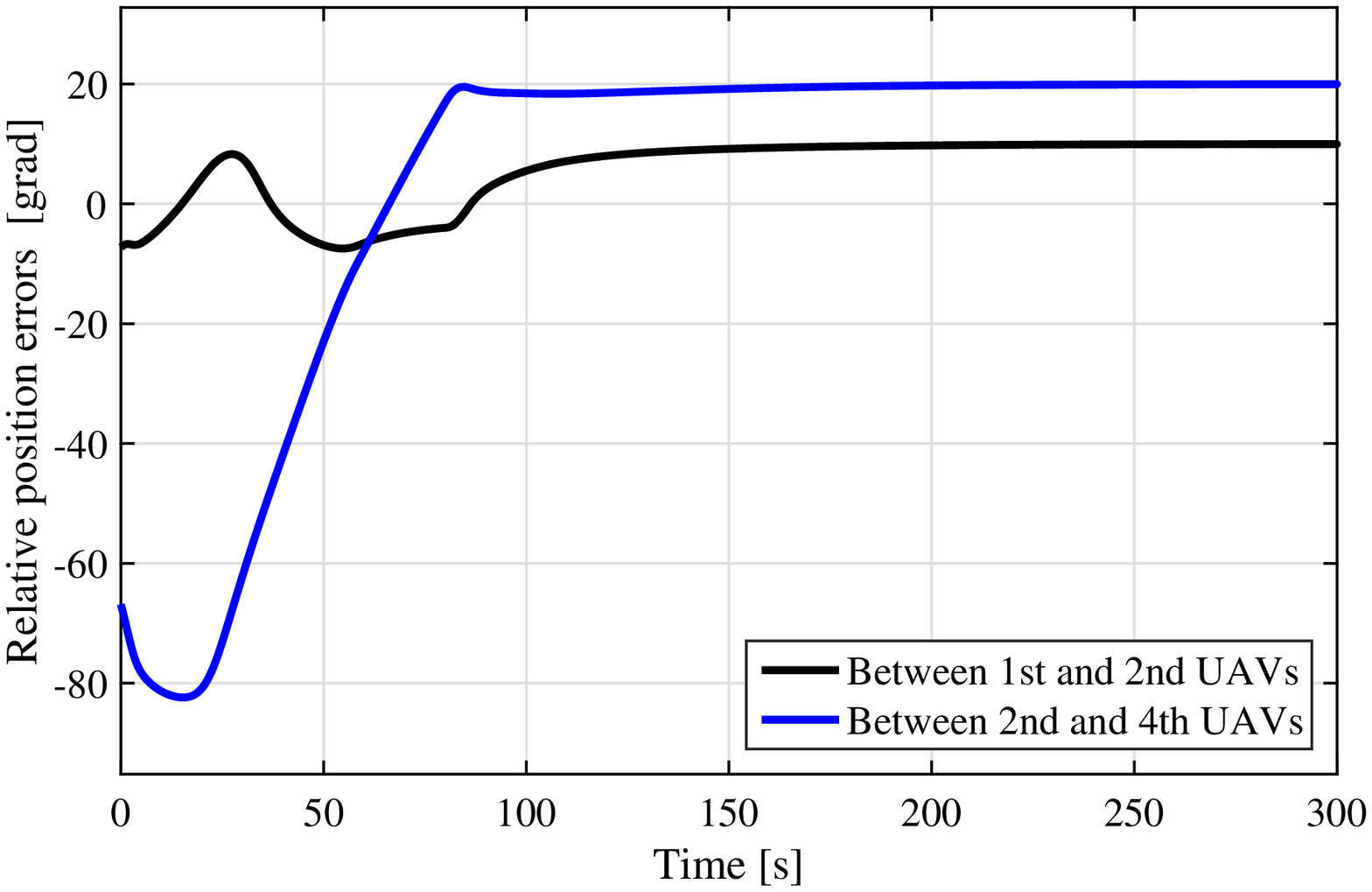}}
	\subfloat[]{\includegraphics[width=0.5\textwidth]{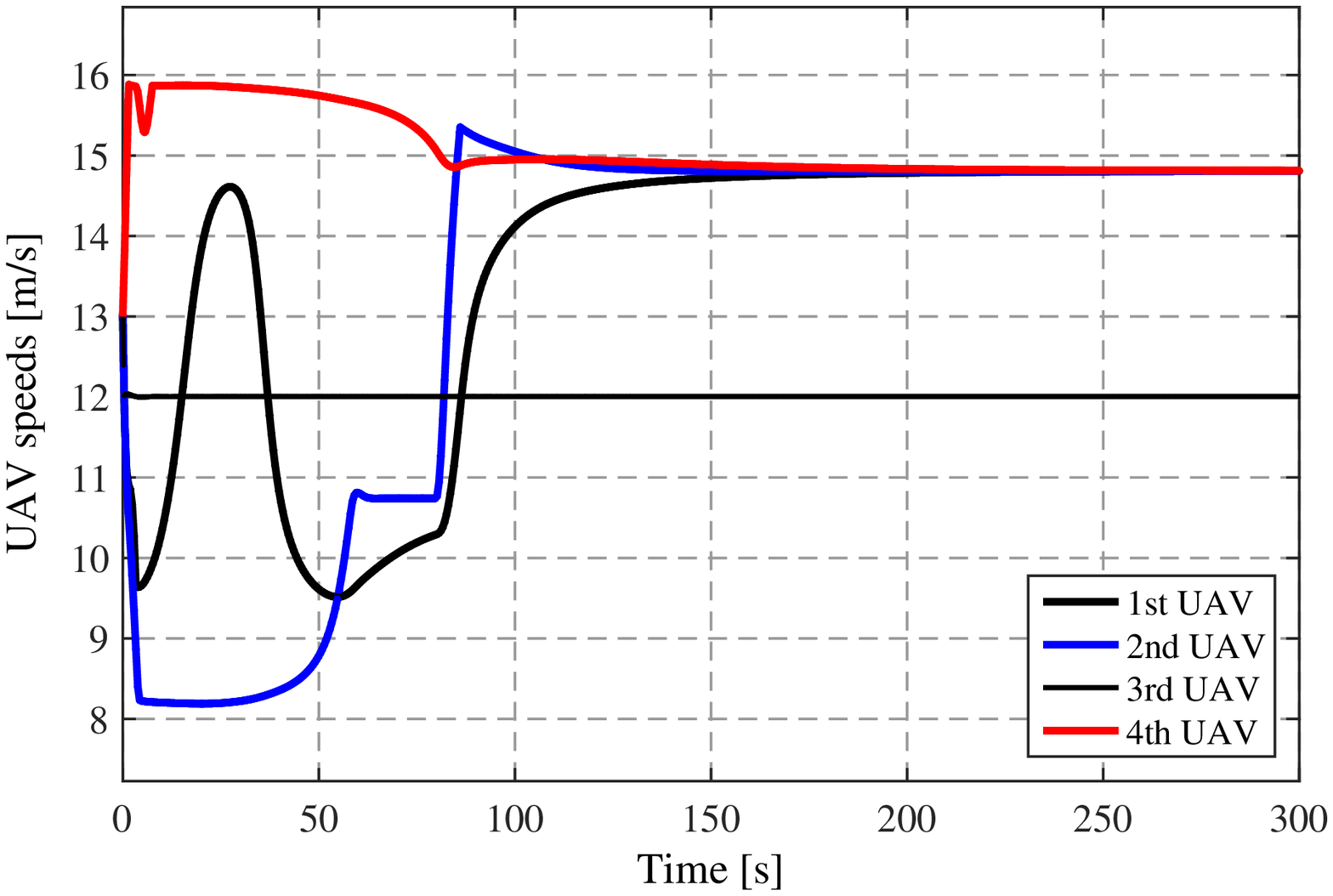}}
	
	\caption{(a) Phase shift errors in the resulting 3-UAV formation after losing 3rd UAV; (b) UAV speeds in the resulting 3-UAV formation after losing 3rd UAV}
\end{figure}

\begin{figure}[b]
	
	\subfloat[]{\includegraphics[width=0.5\textwidth]{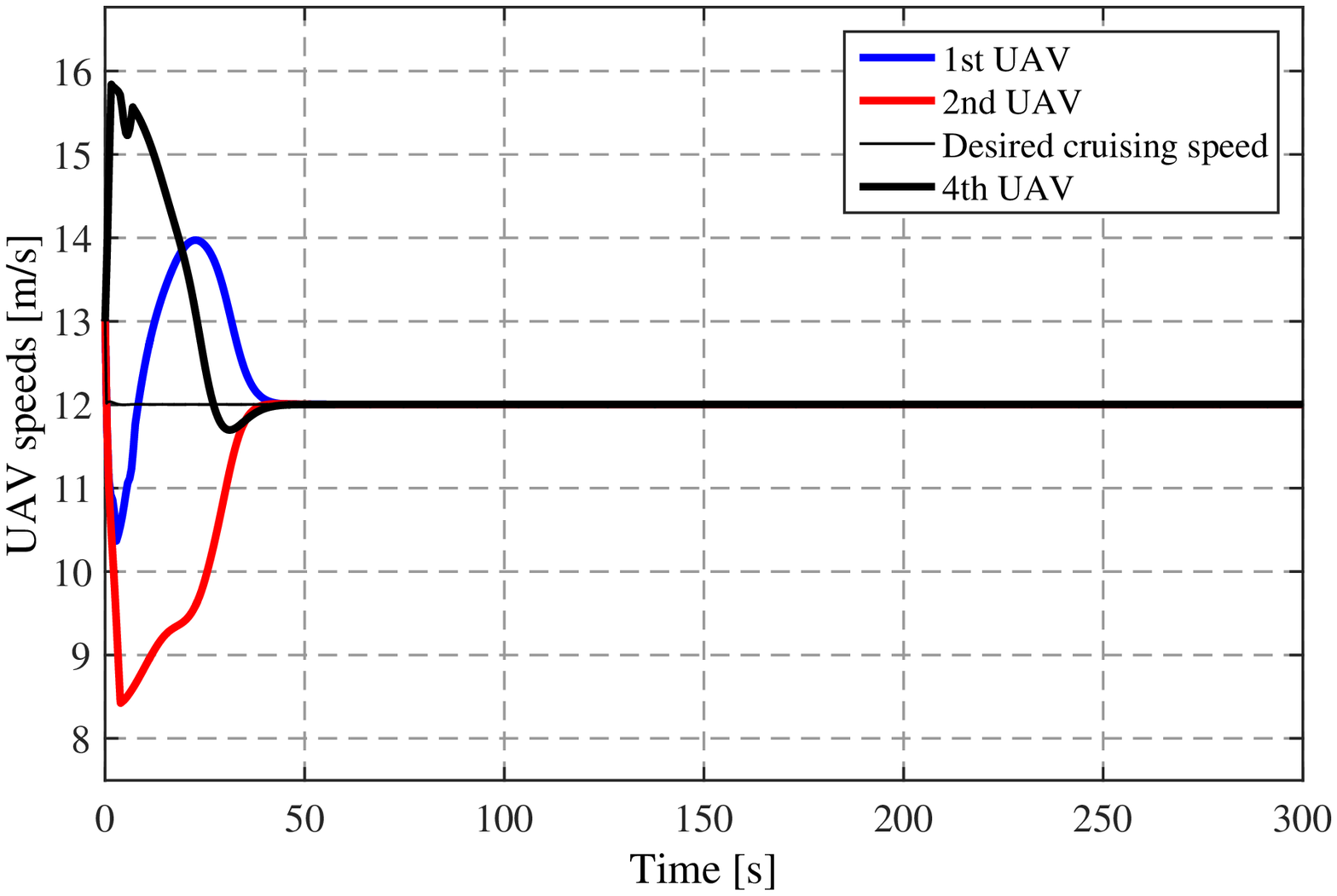}}
	\subfloat[]{\includegraphics[width=0.5\textwidth]{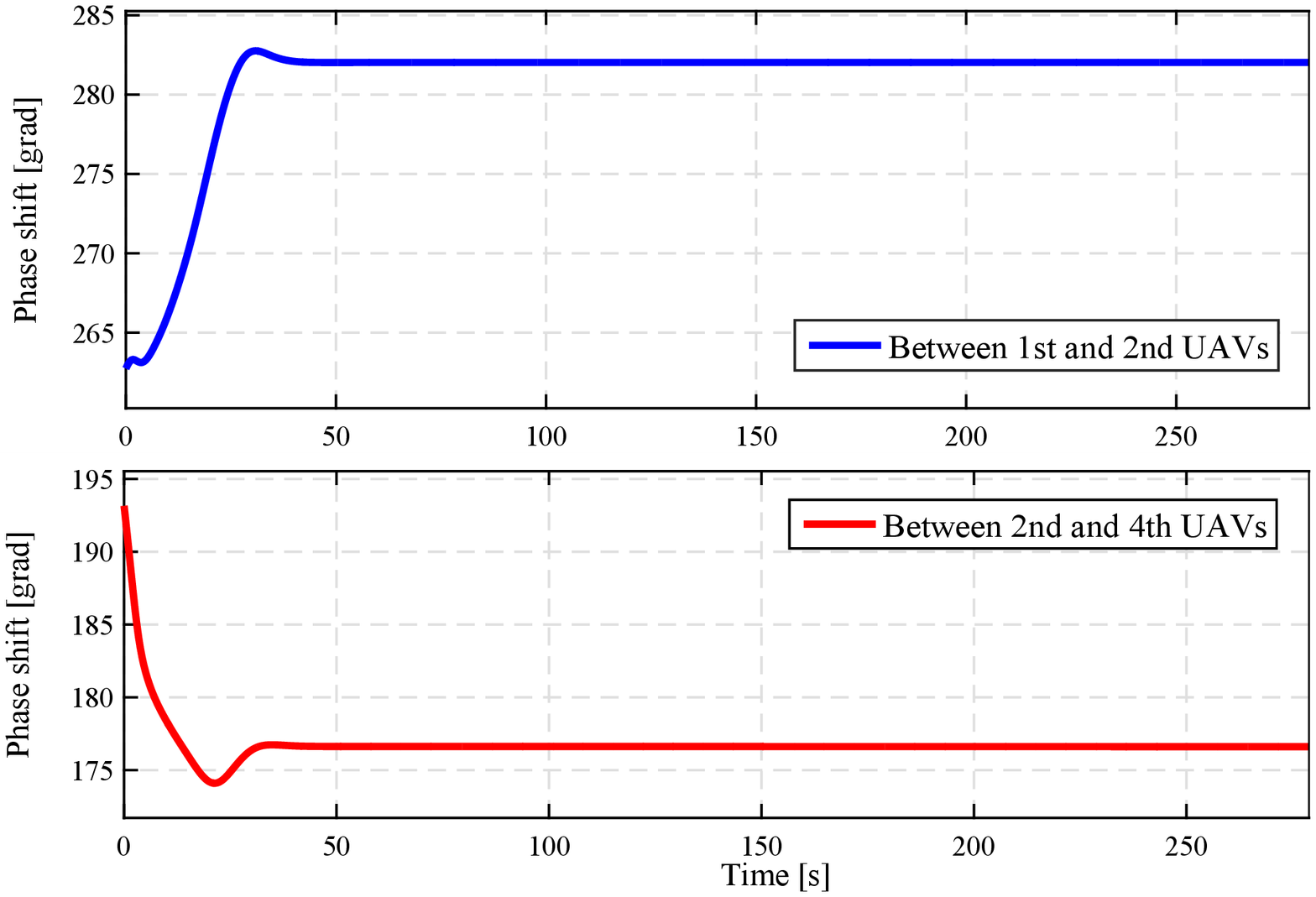}}
	
	\caption{(a) Adaptation algorithm performance in case of adaptively reconfiguring a UAV formation after losing a single UAV; (b) changes in phase shift during post-loss adaptive reconfiguration}
\end{figure}

\section{Simulation Results and Discussion}

A 4-UAV simulation model based on full nonlinear models has been developed in MATLAB/Simulink. For details on the models, refer to the monograph [22]. Simulation parameters are the same as in paper [23]. Find below the results of simulation modeling using full nonlinear models. Fig. 3a shows the trajectories of a 4-UAV formation in case of a stationary target. Fig. 3b shows changes in phase shift errors. Apparently, these errors reach zero over time. There are three phase shift errors as the UAVs in the formation use open-chain interaction. 

Stability can be proved as follows, similarly to [18]. Let the function $E$ (2) be the Lyapunov function. Then derivative of the Lyapunov function $E$ along the trajectories of the system (4), with an arctangent for ${f_{sigm}}$, is written as
\begin{equation*}
\dot E =  - \sum\nolimits_{i = 1}^n {{\tau _p}\frac{2}{\pi }} \left( {{p_i} - {p_{di}}} \right)\arctan \left( {{p_i} - {p_{di}}} \right) < 0,\;\;\forall \left( {{p_i} - {p_{di}}} \right) \ne 0.
\end{equation*}
Here the condition of quasi-stationarity is accepted and taken into account, that is ${p_i} = {\rm{const}}$, and the system manages to attain a new state of equilibrium after losing one or more UAVs, since, as mentioned earlier, the adaptation algorithm has much slower dynamics than the system. 

Fig. 4a shows how phase shift errors change and that they do not reach zero because losing UAV No. 3 makes the pattern unattainable. However, graphs also show the system to reach a new state of equilibrium. There are two phase shift errors here, as the UAVs in the formation use open-chain interaction; losing one makes it a 3-UAV formation. 

As noted earlier, the new equilibrium state exists as a result of interaction balancing. This can be seen in the UAV speeds graph when the formation loses one unit, see Fig. 4b. UAV No. 3 is lost. The remaining UAVs fail to attain the initially preconfigured cruising speed of 12 m/s. These are UAV No. 1 and UAVs No. 2, No. 4 that the lost UAV was between. Failure to reach the required cruising speed is due to the continuous inter-UAV communication that keep the system in this new equilibrium state. Cruising at a speed other than the initially configured value is strongly undesirable for two reasons: (i) it may jeopardize the mission that the initial cruising speed was meant for; (ii) it may be resource-suboptimal as the initial cruising speed was optimized for the UAV's aerodynamics to maximize fuel efficiency. Thus, reducing interaction within the UAV formation subject to adaptive reconfiguration is directly applicable to attaining the initially configured cruising speed.

We further implemented the adaptation algorithm (4) with the coefficient ${\tau _p} = 0.1$. As shown in Fig. 5a, this made every UAV in the formation reach its preconfigured final cruising speed of 12 m/s. Fig. 5b shows change in the phase shift between UAV No. 1 and UAV No. 2, UAV No. 2 and UAV No. 4. The new state of equilibrium is different from that of the 4-UAV formation.

\section{Conclusions}
The adaptation strategy presented herein eliminates a negative effect that completely decentralized UAV formations tend to show when losing one or more UAVs. The formation reconfiguration approach shown here could be applicable in more diverse UAV formation control scenarios; the adaptation algorithms are further modifiable. This adaptation scenario also necessitates designing a diagnostics module for UAVs; further work will present its models. 

\begin{acknowledgement}
	This work was supported by the Ministry of Science and Higher Education of the Russian Federation (Agreement No. 075-15-2021-1016).
\end{acknowledgement}

\end{document}